\documentclass[fleqn,twoside]{article}
\usepackage{espcrc2,cite,graphicx,amsmath,amsfonts,amssymb,mathrsfs,epic,eepic}
\usepackage[figuresright]{rotating}

\newcommand{\ie}{{\it i.e.}}

\newcommand{\eg}{{\it e.g.}}

\newcommand{\cf}{{\it cf.}}
\newcommand{\etc}{{\it etc.}}
\newcommand{\eq}{Eq.}
\newcommand{\eqs}{Eqs.}
\newcommand{\fig}{Fig.}
\newcommand{\tab}{Tab.}
\newcommand{\figs}{Figs.}
\newcommand{\Ref}{Ref.}
\newcommand{\Refs}{Refs.}
\newcommand{\Sec}{Sec.}

\begin{document}

\title{
\vspace*{-3cm}
\begin{flushright}
{\small TUM-HEP-470/02\\[-2.5mm] MPI-PhT/2002-28}
\end{flushright}
\vspace*{0.5cm}
{\bf Tomography of the Earth's Core Using Supernova Neutrinos}}

\author{
{\large Manfred Lindner}\address[TUM]{{\it Institut f{\"u}r Theoretische
Physik, Physik-Department, Technische Universit{\"a}t M{\"u}nchen,
James-Franck-Stra{\ss}e, 85748 Garching bei M{\"u}nchen,
Germany}}\thanks{E-mail: {\tt lindner@ph.tum.de}},
{\large Tommy Ohlsson}\addressmark[TUM]\address{{\it Division of
Mathematical Physics, Department of Physics, Royal Institute of
Technology (KTH) - Stockholm Center for Physics, Astronomy, and
Biotechnology, 106 91 Stockholm, Sweden}}\thanks{E-mail: {\tt
tohlsson@ph.tum.de, tommy@theophys.kth.se}},
{\large Ricard Tom{\`a}s}\address[MPP]{{\it Max-Planck-Institut
f{\"u}r Physik (Werner-Heisenberg-Institut), F{\"o}hringer Ring 6,
80805 M{\"u}nchen, Germany}}\thanks{E-mail:
{\tt ricard@mppmu.mpg.de}},
{\large Walter Winter}\addressmark[TUM]\thanks{E-mail: {\tt
wwinter@ph.tum.de}}
}

\begin{abstract}
\noindent {\bf Abstract}
\vspace{2.5mm}

We investigate the possibility to use the neutrinos coming from a
future galactic supernova explosion to perform neutrino oscillation
tomography of the Earth's core. We propose to use existing or planned
detectors, resulting in an additional payoff. Provided that all of the 
discussed uncertainties can be reduced as expected, we find that the average
matter densities of the Earth's inner and outer cores could be measured with a
precision competitive with geophysics. However, since seismic waves
are more sensitive to matter density jumps than average matter
densities, neutrino physics would give partly complementary information.

\vspace*{0.2cm}
\end{abstract}

\maketitle

\section{Introduction}

In order to obtain more information about the interior of the
Earth, neutrino tomography has been considered as an alternative method to
geophysics. There exist, in principle, two different such techniques, neutrino
absorption
tomography~\cite{Volkova74,DeRujula83,Wilson84,Askar84,Borisov87,Nicolaidis91,Crawford95,Kuo95,Jain:1999kp}
and neutrino oscillation
tomography~\cite{Ermilova:1988pw,Chechin:1991,Ohlsson:2001ck,Ohlsson:2001fy,Ioannisian:2002yj}.
Neutrino absorption tomography, based on the absorption of neutrinos
in matter, is in some sense similar to X-ray
tomography and unfortunately faces several problems including the need
of extremely high energetic neutrino sources, huge detectors, and the
prerequisite of many baselines. Neutrino oscillation
tomography uses the fact that neutrino oscillations are influenced by
the presence of matter~\cite{mikh85,mikh86,wolf78}.
Neutrino oscillation tomography would, in principle, be possible with
a single baseline, since interference effects provide additional
information on the matter density profile. However, it requires quite precise
knowledge about the neutrino oscillation parameters and stringent bounds on the
contribution of non-oscillation physics, such as neutrino decay, CPT
violation, non-standard interactions, sterile neutrinos, \etc{}
Supernovae as neutrino sources are especially
interesting, since the neutrinos come in large numbers from a short 
burst, which could be used to obtain a snapshot of the Earth's interior. In
addition, compared to solar neutrinos, their energy spectrum has a
high-energy tail, which is more sensitive to Earth matter effects. The
influence of Earth matter on supernova neutrinos has, for example, been studied
in \Refs~\cite{Lunardini:2001pb,Takahashi:2001dc,Takahashi:2000it}. 

We assume that technologically feasible detectors exist, such as
Super-Kamiokande, SNO, Hyper-Kamiokande, and UNO, which are originally built
for different purposes, but also capable to detect supernova neutrinos. We
discuss the possibility to use the neutrinos coming from a future
galactic supernova explosion to determine with the assumed detectors some of 
the
measurable quantities describing the structure of the Earth's interior. We
especially focus on the outer and inner core of the Earth,
since they are much harder to access with conventional geophysical methods than
the mantle of the Earth.

\section{Geophysical aspects}

In geophysics, the most promising technique to access the Earth's interior is 
to use seismic wave propagation (for a summary, see \Refs~\cite{aki80,lay95}).
Especially, the detection of seismic waves produced by earthquakes gives
valuable information on the seismic wave velocity profile of the Earth matter. 
However, the
matter density is not directly accessible, but indirectly obtained by
assumptions about the equation of state of the considered
materials. Since seismic S-waves are mainly reflected at the
mantle-core boundary, information on the Earth's core is much harder
to obtain than on the Earth's mantle. Therefore, we will
especially focus on the Earth's core in this paper. Since reflection and
refraction of P-waves at transition boundaries with large matter density 
contrasts are quite easy to observe with seismic waves, the mantle-core and 
outer core-inner core boundaries can be located with high precision as well as 
the matter density jumps
can be measured. For example, the matter density jump at the outer
core-inner core
boundary is often given by $(0.55 \pm 0.05) \,
\mathrm{g/cm^3}$~\cite{Igel}. This is quite a difference compared to neutrino
physics, 
since neutrino oscillations in matter are especially sensitive to the
average matter densities (on the scale of the neutrino oscillation
length) indirectly measured by the electron density. Thus, neutrino
oscillations are less sensitive to local differences, but they involve less
unknowns from the equation of state and could therefore access the
absolute matter densities instead of the matter density jumps.

Several issues regarding the Earth's inner core are considered to be
interesting from a geophysical point of view. For different indirect reasons
the inner core is believed to consist mainly of iron and it is
therefore often called the iron core. First of all, the spectral lines in the
sunlight indicate that the atmosphere of the Sun partly consists of iron as
a potential material source for the planets. A second hint comes from the 
magnetic field of the Earth. After all, there are no convincing
alternatives. We will see later that neutrino tomography could
directly verify the average matter density of an iron core. Further topics
relevant for the inner core structure are: anisotropy, heterogeneity,
time-dependence, solidity, and rotation (for a summary, see
\Ref~\cite{Steinle-Neumann:2002}). However, since neutrino oscillations are 
to a good approximation only sensitive to average matter densities at
long scales, these issues are much harder to access with neutrino
oscillation tomography.

\section{Core-collapse supernovae as neutrino sources}

Core-collapse supernovae represent the evolutionary end of massive
stars with a mass $M \gtrsim 8 M_\odot$, where
$M_\odot$ is the mass of the Sun. In these explosions, about 99~\% of
the liberated gravitational binding energy, $E_b \simeq 3 \cdot
10^{53}$ erg, is carried away by neutrinos in roughly equal amounts of
energy for all flavors in the first 10 seconds after the onset of the
core collapse~\cite{Raffelt:2002tu}. It is widely believed that the
time-dependent energy spectrum of each neutrino species can be approximated by 
a
``pinched'' Fermi--Dirac
distribution~\cite{Janka:1989,Janka:19892,Giovanoni:1989}. In this work, we
assume that the time-integrated spectra can also be well approximated by the
pinched Fermi--Dirac distributions with an effective degeneracy parameter
$\eta$, \ie,
\begin{equation} N_\nu^0(E) = \frac{E_\nu^{\rm tot}}{\langle E_\nu
\rangle} \frac{1}{F_2(\eta) T^3} \frac{E^2}{{\rm e}^{E/T - \eta} + 1},
\label{eq:FD}
\end{equation}
where $E_\nu^{\rm tot} \equiv \int L_\nu \, dt$ is the total neutrino energy 
of a certain flavor $\nu$,
$$
F_k(y) \equiv \int_0^\infty \frac{x^k}{{\rm e}^{x-y} + 1} \, dx,
$$
and $E$ and $T$ denote the neutrino energy and the effective
temperature, respectively. The relation between the average
neutrino energy and the effective neutrino temperature is given by
\begin{equation}
\langle E_\nu \rangle = \frac{\int_0^\infty E N_\nu^0 \,
  dE}{\int_0^\infty N_\nu^0 \, dE} = \frac{F_3(\eta)}{F_2(\eta)} T.
\end{equation}
In this section, we assume for simplicity that $\eta=0$ for all 
neutrino flavors\footnote[1]{In the analysis, we will also show some 
results for $\eta_{\bar{\nu}_e}=3$ and $\eta_{\bar{\nu}_\mu}=1$
\cite{Keil:2002in}. The extension of the formulas to these cases are
quite straightforward, though making them more complicated.}, which means that
\begin{equation}
\langle E_\nu \rangle = \frac{F_3(0)}{F_2(0)} T = \frac{7 \pi^4}{180
  \zeta(3)} T \equiv k T,
\end{equation}
where $\zeta(x)$ is the Riemann z-function [$\zeta(3) \simeq 1.20206$]
and $k \simeq 3.15137$. Furthermore, the time-integrated flux of the
neutrinos can be expressed by
\begin{equation}
\Phi_\nu^0 = \frac{N_\nu^0}{4 \pi D^2},
\end{equation}
where $D$ is the distance to the supernova.

Due to the different trapping processes, the different neutrino
flavors originate in layers of the supernova with different
temperatures. The electron (anti)neutrino flavor is kept in thermal
equilibrium by $\beta$ processes up to a certain radius usually referred to as
the ``neutrinosphere'', beyond which the neutrinos stream off
freely. However, the practical absence of muons and taus in the
supernova core implies that the other two neutrino flavors, here
collectively denoted by $\nu_x$ ($\nu_\mu, \nu_\tau, \bar\nu_\mu,
\bar\nu_\tau$), interact primarily by less efficient neutral-current processes.
Therefore, their spectra are determined at deeper, \ie, hotter,
regions. In addition, since the content of neutrons is larger
than that of protons, $\nu_e$'s escape from outer regions than
$\bar{\nu}_e$'s. This rough picture leads to the following hierarchy:
$\langle E_{\nu_e}\rangle < \langle E_{\bar\nu_e}\rangle < \langle
 E_{\nu_x}\rangle $.
Here $\nu_x$ refers again to both $\nu_\mu$ and $\nu_\tau$.
Typical values of the average energies of the time-integrated neutrino
spectra obtained in simulations
are $\langle E_{{\nu}_e} \rangle \sim 12$ MeV, $\langle E_{\bar{\nu}_e} \rangle
\sim 15$ MeV, and $\langle E_{{\nu}_x} \rangle \sim 24$ MeV
\cite{Janka:1992jk,Totani:1998vj}.
However, recent studies with an improved treatment of neutrino transport,
microphysics, the inclusion of the nucleon bremsstrahlung, and the
energy transfer by recoils, find somewhat smaller differences between the
$\bar\nu_e$ and $\nu_x$ spectra~\cite{Raffelt:2001ai,Buras:2002wt}.

In the following, we assume a future galactic supernova explosion at a typical
distance of $D = 10$~kpc, with a binding energy of $E_b = 3 \cdot
10^{53}$~erg and a total energy of $E^{\rm tot}_{\nu_e} = E^{\rm
tot}_{\bar\nu_e} \equiv E^{\rm tot}_{\nu_x}/\xi$, where $\xi$
parameterizes a possible deviation from energy
equipartition~\cite{Mezzacappa:2000jb}. We also assume that the fluxes of
$\nu_\mu$, $\nu_\tau$, $\bar\nu_\mu$, and $\bar\nu_\tau$ are identical and we
fix $\langle E_{\bar{\nu}_e} \rangle$ to 15~MeV.
Due to the current lack of a standard picture of core-collapse
supernovae we consider for our analysis six scenarios for $\eta=0$
with different values of the parameters $\tau_E \equiv \langle
E_{\bar{\nu}_x} \rangle/ \langle E_{\bar{\nu}_e} \rangle \in
\{1.2,~1.4\}$ and $\xi \in \{0.5,~1,~1.2\}$, as shown in
\tab~\ref{snscenarios}. In addition, we study four scenarios with
$\eta_{\bar{\nu}_e}=3$ and $\eta_{\bar{\nu}_\mu}=1$ as well as $\tau_E
\in \{1.2,~1.4\}$ and $\xi \in \{1,~1.2\}$.

As far as the neutrino detection is concerned, we only analyze the
charged-current reaction
$\bar\nu_e + p \rightarrow e^+ + n$, since this reaction
yields the largest number of events (around 8000 in the
Super-Kamiokande detector in the case of a galactic supernova).
Therefore, we shall concentrate on the study of the propagation of
antineutrinos from a supernova to detectors on the Earth. The cross
section details for the reaction $\bar\nu_e + p \rightarrow e^+ + n$
can be found and are discussed in \Ref~\cite{Beacom:1998fj}.

\begin{table}
\begin{center}
\begin{tabular}{lcccc}
 \hline
 Scenario & $\xi$ & $\tau_E$ & $\eta_{\bar{\nu}_e}$ &  
$\eta_{\bar{\nu}_\mu}$  \\
 \hline
S1 & 1 & 1.4 & 0 & 0 \\
S2 & 1 & 1.2 & 0 & 0 \\
S3 & 0.5 & 1.4 & 0 & 0 \\
S4 & 0.5 & 1.2 & 0 & 0 \\
S5 & 1.2 & 1.4 & 0 & 0 \\
S6 & 1.2 & 1.2 & 0 & 0 \\
S7 & 1 & 1.4 & 3 & 1 \\
S8 & 1 & 1.2 & 3 & 1 \\
S9 & 1.2 & 1.4 & 3 & 1 \\
S10 & 1.2 & 1.2 & 3 & 1 \\  \hline
\end{tabular}
\end{center}
\caption{Our standard scenarios for
supernova parameters, where in all cases
$ \langle E_{\bar\nu_e} \rangle = 15 \, \mathrm{MeV}$.}
\label{snscenarios}
\end{table}


\section{From neutrino production to neutrino detection}
\label{Sec:earth}

In general, neutrino propagation from a source to a detector is
described by an evolution operator on the form
\begin{equation}
\mathcal{U} \equiv \mathcal{U}(L) = {\rm e}^{-{\rm i} \int_0^L
  \mathscr{H}(L') \, dL'},
\end{equation}
where the exponential function is time-ordered, $\mathscr{H} \equiv
\mathscr{H}(L)$ is the total Hamiltonian and $L$ is the neutrino
(traveling) path length, \ie, the baseline 
length. The Hamiltonian is usually given either in the flavor basis or
in the mass basis. In the flavor basis, the total Hamiltonian reads 
$\mathscr{H}_f(L) = U H_m U^{-1} + A(L) \, {\rm diag \,}(1,0,0)$,
where $H_m \equiv {\rm diag \,}(E_1,E_2,E_3)$ is the free Hamiltonian in
the mass basis, $U \equiv (U_{\alpha a})$ is the leptonic mixing matrix,
and
\begin{equation}
A \equiv A(L) = \pm \sqrt{2} G_{\rm F} \frac{Y_e}{m_N} \rho(L)
\label{eq:A}
\end{equation}
is the mass density parameter related to the matter density $\rho \equiv
\rho(L)$. Here $E_a \equiv m_a^2/(2 E)$ ($a = 1,2,3$), $G_{\rm F} \simeq
1.16639 \cdot 10^{-23} \, {\rm eV}^{-2}$ is the Fermi weak coupling
constant, $Y_e$ is the average number of electrons per
nucleon\footnote[2]{In the Earth: $Y_e \simeq \tfrac{1}{2}$.}, and $m_N
\simeq 939 \, {\rm MeV}$ is the nucleon mass. The sign depends on
the presence of neutrinos ($+$) or antineutrinos ($-$).
Furthermore, $m_a$ is
the mass of the $a$th mass eigenstate $\nu_a$ and $E$ is the neutrino energy.
In order to obtain the neutrino oscillation transition probabilities,
we need to calculate the matrix elements of the evolution operator in the
flavor basis, take the absolute values of these, and then square
them. The neutrino oscillation probability amplitude from a neutrino
flavor $\nu_\alpha$ to a neutrino flavor $\nu_\beta$ is defined as
\begin{equation}
A_{\alpha\beta} \equiv \langle \nu_\beta | \mathcal{U}_f(L) | \nu_\alpha
\rangle, \quad \alpha,\beta = e,\mu,\tau,
\end{equation}
where $\mathcal{U}_f$ is the total evolution operator in the flavor
basis. Then, the neutrino oscillation transition probability for $\nu_\alpha
\rightarrow \nu_\beta$ is given by
\begin{equation}
P_{\alpha\beta} \equiv |A_{\alpha\beta}|^2, \quad \alpha,\beta =
e,\mu,\tau.
\end{equation}

The initial neutrino fluxes arise from the central
part of the supernova, where the
matter density is of the order of about $10^{12}~{\rm g/cm}^3$.
For such high matter densities one can infer from the expression of the 
Hamiltonian
$\mathscr{H}_f$ that the matter mass eigenstates, $\nu_a^m$ ($a =
1,2,3$), coincide with the flavor states, $\nu_\alpha$ ($\alpha =
e,\mu,\tau$), up to a rotation between $\nu_\mu$ and $\nu_\tau$.
Thus, in the case of normal mass hierarchy, $m_1 \lesssim m_2 \ll m_3$, one has
\begin{equation}
\bar\nu_1^m = \bar\nu_e, \quad \bar\nu_2^m = \bar\nu_\mu', \quad
\bar\nu_3^m = \bar\nu_\tau',
\end{equation}
where $\bar\nu_\mu'$ and $\bar\nu_\tau'$ are the rotated states.
Therefore, one can assume that the original fluxes of the matter mass
eigenstates are\footnote[3]{Any rotation between $\bar\nu_\mu$
  and $\bar\nu_\tau$ does not affect the corresponding total mass
  eigenstate contents, because they have the same fluxes, as discussed in the 
  last section. For an analysis taking into account possible differences
  in the fluxes, see \Ref~\cite{Akhmedov:2002zj}.
  In addition, an argument against neutrino oscillations between
  $\bar\nu_\mu$ and $\bar\nu_\tau$ on their way to the Earth will be
  given at the end of this section.}
\begin{equation}
\Phi_{\bar{\nu}_1^m}^0 = \Phi_{\bar\nu_e}^0, \quad \Phi_{\bar{\nu}_2^m}^0 =
\Phi_{\nu_x}^0, \quad \Phi_{\bar{\nu}_3^m}^0 = \Phi_{\nu_x}^0.
\label{eq:corrfluxes}
\end{equation}
Since we are assuming that $\bar\nu_\mu$ and $\bar\nu_\tau$ have the same
fluxes, the neutrino transitions are determined by the mixings of the
$\bar\nu_e$ only, \ie, by $U_{ea}$~\cite{Dighe:1999bi}.
Moreover, under the assumption of normal mass hierarchy, antineutrinos
do not undergo any resonant conversion, which means that the small mixing angle
$\theta_{13}$ is suppressed in matter and the
$\bar\nu_e \leftrightarrow \bar\nu_3$ transitions are negligible.
One consequence is that $\bar\nu_3^m$ propagates
adiabatically and leaves the supernova as $\bar\nu_3$.
The propagation of the other two states depends on the parameters of
the solution to the solar neutrino problem and it may be adiabatic or
non-adiabatic~\cite{Smirnov:1994ku,Kachelriess:2001sg}.
In particular, we will focus on the Mikheyev--Smirnov--Wolfenstein
(MSW) \cite{mikh85,mikh86,wolf78} large mixing angle (LMA) solution,
since it is by far the most favored one.
For the parameters within such a region the neutrino evolution through
the supernova (SN) envelope is adiabatic. Thus, $\bar\nu_e$ will leave
the supernova as
$\bar\nu_1$, $\bar\nu_\mu'$ as $\bar\nu_2$, and $\bar\nu_\tau'$ as
$\bar\nu_3$.
Finally, the measured fluxes of supernova neutrinos at a detector on the
Earth are
\begin{equation}
\Phi_{\nu_\alpha} = \sum_{a=1}^3 P_{a \alpha} \Phi^0_{\nu_a}, \quad
\alpha = e,\mu,\tau,
\label{eq:fluxes}
\end{equation}
where $\Phi^0_{\nu_a}$ ($a = 1,2,3$) are the initial supernova
neutrino fluxes.

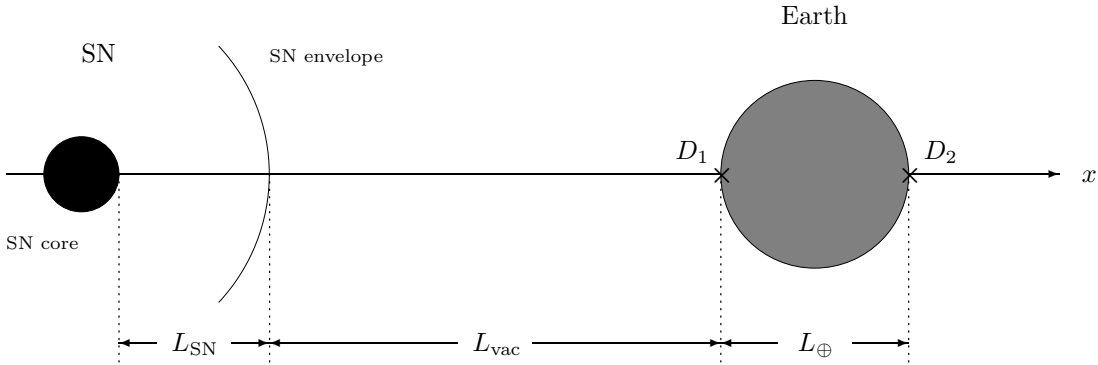
\begin{figure*}
\begin{picture}(16,5)
\put(1,2.5){\vector(1,0){14}} \put(15.3,2.4){$x$}
\put(2,2.5){\blacken\circle{1}} \put(2,2.5){\arc{5}{-0.75}{0.75}}
\put(11.75,2.5){\shade\circle{2.5}}
\put(10.36,2.4){$\mbox{\boldmath{$\times$}}$} \put(9.9,2.7){$D_1$}
\put(12.86,2.4){$\mbox{\boldmath{$\times$}}$} \put(13.2,2.7){$D_2$}
\dottedline{0.1}(2.5,2.5)(2.5,0)
\dottedline{0.1}(4.5,2.5)(4.5,0)
\dottedline{0.1}(10.5,2.5)(10.5,0)
\dottedline{0.1}(13,2.5)(13,0)
\put(3,0.25){\vector(-1,0){0.5}}
\put(4,0.25){\vector(1,0){0.5}}
\put(3.2,0.15){$L_{\rm SN}$}
\put(7,0.25){\vector(-1,0){2.5}}
\put(8,0.25){\vector(1,0){2.5}}
\put(7.2,0.15){$L_{\rm vac}$}
\put(11.25,0.25){\vector(-1,0){0.75}}
\put(12.25,0.25){\vector(1,0){0.75}}
\put(11.525,0.15){$L_\oplus$}
\put(2,4){SN}
\put(11.3,4.5){Earth}
\put(1,1.5){\scriptsize SN core}
\put(4.5,4){\scriptsize SN envelope}
\end{picture}
\caption{The propagation of neutrinos from a supernova (SN) to the
  Earth. The detector $D_1$ can be anywhere on the Earth's side towards the
SN, whereas the detector $D_2$ should be in the shadow of the Earth's
core.} \label{fig:SN-E} \end{figure*}
\begin{figure}[ht!]
\begin{center}
\includegraphics*[width=7cm]{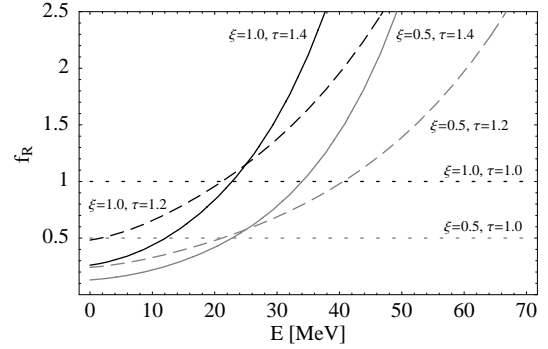}
\end{center}
\vspace{-1cm}
\caption{The flux ratio $f_R$ as a function of the neutrino energy $E$
  for different values of the parameters $\xi$ and $\tau_E$ and for
  $\eta = 0$, partly corresponding to the scenarios in \tab~\ref{snscenarios}.}
\label{fig:fr}
\end{figure}

For neutrino propagation from a supernova (SN core) to a detector at the Earth
(see \fig~\ref{fig:SN-E})
we have the probability amplitudes
\begin{eqnarray}
A_{\alpha\beta} &=& \langle \nu_\beta | \mathcal{U}^{\rm tot}_f(L) | \nu_\alpha
\rangle \nonumber\\
&=& \langle \nu_\beta | \mathcal{U}^\oplus(L_\oplus) \mathcal{U}^{\rm
  vac}(L_{\rm vac}) \mathcal{U}^{\rm SN}(L_{\rm SN}) | \nu_\alpha
\rangle,
\nonumber
\\
\label{eq:totalop}
\end{eqnarray}
where $\mathcal{U}^{\rm SN}$, $\mathcal{U}^{\rm vac}$, and
$\mathcal{U}^\oplus$ are the evolution operators in the supernova
(from SN core to SN envelope), in vacuum, and in the Earth, respectively, and
$L_{\rm SN}$, $L_{\rm vac}$, and $L_\oplus$ are the corresponding
baseline lengths. Note that the operators $\mathcal{U}^{\rm SN}$,
$\mathcal{U}^{\rm vac}$, and $\mathcal{U}^\oplus$ in general do not commute.
Using the completeness relation, one can write the probability
amplitudes as
\begin{eqnarray}
A_{\alpha\beta} & = & \sum\limits_{a=1}^3 \langle \nu_\beta |
\mathcal{U}^\oplus(L_\oplus)
  \mathcal{U}^{\rm vac}(L_{\rm vac}) | \nu_a \rangle  \times \nonumber
  \newline \\
 & & \langle \nu_a |
  \mathcal{U}^{\rm SN}(L_{\rm SN}) | \nu_\alpha \rangle.
 \label{eq:atotal}
\end{eqnarray}
Since we have seen that for adiabatic transitions the supernova
neutrinos leave the supernova (SN envelope) as neutrino mass
eigenstates $\nu_a$, \ie, 
$\langle \nu_a |  \mathcal{U}^{\rm SN}(L_{\rm SN}) | \nu_\alpha \rangle=
  \delta_{\alpha a}$, we can re-define the probability amplitudes
\begin{equation}
A_{a \alpha} \equiv \langle \nu_\alpha | \mathcal{U}^\oplus (L_E)
\mathcal{U}^{\rm vac}(L_{\rm vac}) | \nu_a \rangle,
\end{equation}
where the first index is a mass eigenstate index ($a = 1,2,3$) and the
second index is a flavor state index ($\alpha = e,\mu,\tau$). These
``mixed'' probability amplitudes will completely determine the
evolution of the neutrinos from a supernova (SN envelope) to the Earth.
Now, there are, in principle, two cases for a supernova neutrino to be
detected at the Earth (see again \fig~\ref{fig:SN-E}):
\begin{enumerate}
\item The supernova neutrino arrives at the detector from above, \ie,
  it does not go through the Earth at all (detector $D_1$).
\item The supernova neutrino goes through the Earth's and then arrives at
  the detector (detector $D_2$).
\end{enumerate}
Let us start with the first case. The probability amplitude for an
initial neutrino mass eigenstate $\nu_a$, where $a = 1,2,3$, to leave
the supernova and to convert into a flavor state $\nu_\alpha$, where
$\alpha = e,\mu,\tau$, is at the detector $D_1$ formally defined as
\begin{equation}
A^{D_1}_{a \alpha} = \langle \nu_\alpha | \mathcal{U}^{\rm vac}(L_{\rm
  vac}) | \nu_a \rangle.
\label{eq:AD1}
\end{equation}
Note that we assumed that a neutrino mass eigenstate $\nu_a$ left the
supernova, and therefore, no evolution operator $\mathcal{U}^{\rm SN}$ should
appear in the above equation. Furthermore, since the supernova
neutrino does not go through the Earth, there appears also no
evolution operator $\mathcal{U}^\oplus$ in this equation. Next, since the
evolution operator in vacuum $\mathcal{U}^{\rm
vac}$ is diagonal in the mass basis, we find that \eq~(\ref{eq:AD1})
reduces to\footnote[4]{The neutrino flavor states are defined as follows:
  $| \nu_\alpha \rangle = \sum_{a=1}^3 U_{\alpha a}^\ast | \nu_a
  \rangle$ ($\alpha = e,\mu,\tau$), which implies that $\langle
  \nu_\alpha | = \sum_{a=1}^3 U_{\alpha a} \langle \nu_a |$. Here the
  $U_{\alpha a}$'s are the matrix elements of the leptonic mixing
  matrix $U$.}
\begin{eqnarray}
A^{D_1}_{a \alpha} &=& \langle \nu_\alpha | \nu_a \rangle =
  \sum_{b=1}^3 U_{\alpha b} \langle \nu_b | \nu_a \rangle \nonumber\\
&=& \sum_{b=1}^3 U_{\alpha b} \delta_{ab} = U_{\alpha a},
\end{eqnarray}
\ie, the probability amplitudes are just the matrix elements of the
leptonic mixing matrix.
Thus, we have $P^{D_1}_{a \alpha} = | A^{D_1}_{a \alpha} |^2 =
| U_{\alpha a} |^2$, and inserting this into \eq~(\ref{eq:fluxes}),
we obtain the supernova neutrino fluxes at the detector $D_1$ as
\begin{equation}
\Phi^{D_1}_{\nu_\alpha} = \sum_{a=1}^3 | U_{\alpha a} |^2
\Phi^0_{\nu_a}, \quad \alpha = e,\mu,\tau.
\label{eq:fluxD1}
\end{equation}
Assuming as in \eq~(\ref{eq:corrfluxes}) that the initial
fluxes of the second and third mass
eigenstates are equal, \ie, $\Phi^0_{\bar\nu_2} = \Phi^0_{\bar\nu_3}$, the
electron antineutrino flux at the detector $D_1$
can be written as
\begin{equation}
\Phi^{D_1}_{\bar\nu_e} = \Phi^0_{\bar\nu_1} \left[ | U_{e1} |^2 + f_R \left(
  | U_{e2} |^2 + | U_{e3} |^2 \right) \right].
\label{eq:fluxD12}
\end{equation}
Here $f_R \equiv \Phi^0_{\bar\nu_2}/\Phi^0_{\bar\nu_1} =
\Phi^0_{\bar\nu_3}/\Phi^0_{\bar\nu_1}$ is the so-called flux
ratio, which is plotted for several values of $\xi$ and $\tau_E$
(introduced in the last section) in \fig~\ref{fig:fr}.
Furthermore, the flux ratio $f_R$ depends on the supernova parameters
$\langle E_{\bar{\nu}_e} \rangle$, $\xi$, and $\tau_E$ only and reads for
$\eta = 0$~\footnote[5]{For $\eta \neq 0$ this equation would 
be slightly more complicated, but could be easily obtained from
\eq~(\ref{eq:FD}).}
\begin{equation}
f_R = \frac{\xi}{\tau_E^4} \frac{{\rm e}^{k \frac{E}{\langle E_{\bar{\nu}_e}
      \rangle}} +1}{{\rm e}^{k \frac{E}{\langle E_{\bar{\nu}_e}
      \rangle} \tau_E^{-1}} +1},
\label{eq:deffr}
\end{equation}
where again $k \simeq 3.15137$.
Reinserting the probabilities $P^{D_1}_{ae}$ instead of the
probability amplitudes $U_{ea}$ in \eq~(\ref{eq:fluxD12}) and
using the unitarity condition $P^{D_1}_{1e} + P^{D_1}_{2e} + P^{D_1}_{3e} = 1$,
we find that
\begin{equation}
\Phi^{D_1}_{\bar\nu_e} = \Phi^0_{\bar\nu_1} P^{D_1}_{1e} \left( 1 +
  f_R \frac{1 - P^{D_1}_{1e}}{P^{D_1}_{1e}} \right),
  \label{eq:fluxd1}
\end{equation}
which means that the flux of electron antineutrinos at the detector $D_1$
is only depending on the initial neutrino flux $\Phi^0_{\bar\nu_1}$, the
transition probability $P^{D_1}_{1e} = | U_{e1} |^2$, and the
flux ratio $f_R$.

Next, let us discuss the second case. Again using the fact that the
evolution operator in vacuum $\mathcal{U}^{\rm vac}$ is diagonal in the mass
basis, we find that
\begin{eqnarray}
A^{D_2}_{a \alpha} &=& \langle \nu_\alpha |
\mathcal{U}^\oplus(L_\oplus) \mathcal{U}^{\rm vac}(L_{\rm vac}) |
\nu_a \rangle \nonumber\\
&=& \sum_{b=1}^3 \langle \nu_\alpha | \mathcal{U}^\oplus(L_\oplus) | \nu_b
  \rangle \langle \nu_b | \mathcal{U}^{\rm vac}(L_{\rm vac}) | \nu_a
  \rangle \nonumber\\
&=& \sum_{b=1}^3 \langle \nu_\alpha | \mathcal{U}^\oplus(L_\oplus) | \nu_b
  \rangle \delta_{ab} \nonumber\\
&=& \langle \nu_\alpha | \mathcal{U}^\oplus(L_\oplus) | \nu_a \rangle =
  A^\oplus_{a \alpha}. 
\label{eq:AD2}
\end{eqnarray}
Similar to the first case, we obtain the supernova neutrino
fluxes at the detector $D_2$ as
\begin{equation}
\Phi^{D_2}_{\nu_\alpha} = \sum_{a=1}^3 | A^\oplus_{a \alpha} |^2
\Phi^0_{\nu_a}, \quad \alpha = e,\mu,\tau,
\label{eq:fluxD2}
\end{equation}
which for the electron antineutrino flux at the detector $D_2$ can be
written as
\begin{equation}
\Phi^{D_2}_{\bar\nu_e} = \Phi^0_{\bar\nu_1} P^{D_2}_{1e} \left( 1 +
  f_R \frac{1 - P^{D_2}_{1e}}{P^{D_2}_{1e}} \right).
  \label{eq:fluxd2}
\end{equation}
This means that the flux of electron antineutrinos at the detector $D_2$
depends only on the initial neutrino flux $\Phi^0_{\bar\nu_1}$, the
transition probability $P^{D_2}_{1e} = | A^{D_2}_{1e} |^2$, and the
flux ratio $f_R$.

Now, we want to determine the neutrino oscillation transition probabilities.
Using the evolution operator method developed in \Ref\cite{Ohlsson:1999um}, the
evolution operator in the Earth, which we will assume to consist of $N$
different (constant) matter density layers, is given by
\begin{eqnarray}
\mathcal{U}^\oplus(L_\oplus) &=& \mathcal{U}^\oplus(L_N;A_N)
\mathcal{U}^\oplus(L_{N-1};A_{N-1}) \ldots \nonumber\\
&\times& \mathcal{U}^\oplus(L_2;A_2) \mathcal{U}^\oplus(L_1;A_1) \nonumber\\
&\equiv& \prod_{k=1}^N \mathcal{U}^\oplus(L_k;A_k),
\label{eq:evolE}
\end{eqnarray}
where $\mathcal{U}^\oplus(L_k;A_k) \equiv {\rm e}^{-i \mathscr{H}(A_k)
L_k}$ is the evolution operator in the $k$th layer with constant
matter density and $L_\oplus \equiv \sum_{k=1}^N
L_k$.\footnote[6]{Similar applications of the
evolution operator for propagation of neutrinos in matter consisting
  of two density layers using two neutrino flavors have been discussed
  in \Refs\cite{Akhmedov:1988kd,Petcov:1998su,Akhmedov:1998ui}.}
Here $L_k$ and $A_k$ are the thickness and matter density parameters of
the $k$th matter density layer, respectively. Note again that the evolution
operators in the different layers normally do not commute.
Inserting \eq~(\ref{eq:evolE}) into \eq~(\ref{eq:AD2}) and the result
thereof into $P^{D_2}_{a\alpha} = |A^{D_2}_{a\alpha}|^2$, we finally obtain
\begin{equation}
P^{D_2}_{a\alpha} = \left| \langle \nu_\alpha |
\prod_{k=1}^N \mathcal{U}^\oplus(L_k;A_k) | \nu_a \rangle \right|^2,
\label{eq:PD2_E}
\end{equation}
which is our main formula for the neutrino oscillation transition
probabilities from a supernova to the detector $D_2$. Thus, inserting $a =
1$ and $\alpha = e$ into \eq~(\ref{eq:PD2_E}), we find the probability
$P^{D_2}_{1e}$, which can then be inserted into \eq~(\ref{eq:fluxd2}).

\begin{figure}[ht!]
\begin{center}
\includegraphics*[width=7cm]{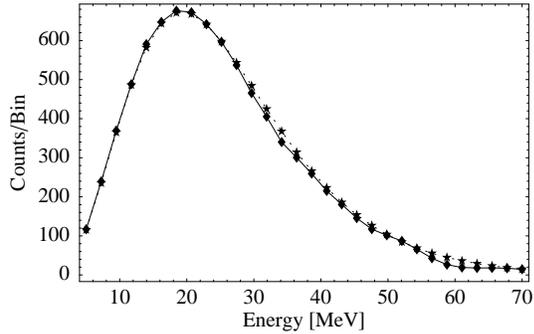}
\end{center}
\vspace{-1cm}
\caption{\label{Rates} The charged-current events per energy bin from
electron antineutrinos in an Super-Kamiokande-like detector for a
realistic energy resolution~\cite{Bahcall:1997ha}. The solid curve
shows the spectrum (scenario S1 in \tab~\ref{snscenarios}) including
the Earth matter effects for a baseline of about $12{,}700 \,
\mathrm{km}$ through the Earth's core, whereas the dashed curve shows
the spectrum without Earth matter effects. The differences appear to
be small, but are nevertheless statistically significant, \ie, $\Delta
\chi^2 \simeq 35$. Assuming that the matter effects can
(approximately) be described by two parameters, such as the average
mantle and core densities, this corresponds to at least a $5 \sigma$
effect. To establish the Earth matter effects at a $3 \sigma$
confidence level, one would need at least $11 \, \mathrm{kt}$ fiducial
mass detectors.
For the neutrino
oscillation parameters we used $\theta_{12}=32.9^\circ$, $\theta_{13}=5^\circ$,
$\theta_{23}=45^\circ$, $\Delta m_{21}^2 = 5.0 \cdot 10^{-5} \,
\mathrm{eV}^2$, $\Delta m_{32}^2 = 2.5 \cdot 10^{-3} \,
\mathrm{eV}^2$, and $\delta_{\mathrm{CP}}=0$, \ie, a normal mass hierarchy.}
\end{figure}

In \fig~\ref{Rates}, we show for the scenario S1 in
\tab~\ref{snscenarios} the different binned energy spectra at
Super-Kamiokande-like detectors, where the solid curve represents the
detector $D_2$ and the dashed curve the detector $D_1$. One can
easily see that matter effects are largest for energies above about
$25 \, \mathrm{MeV}$. For a detailed discussion of Earth matter
effects of supernova neutrinos, see
\Refs~\cite{Lunardini:2001pb,Takahashi:2001dc,Takahashi:2000it} and
references therein. In summary, for our chosen values of the neutrino
oscillation parameters (MSW LMA
solution, maximal atmospheric mixing, and normal mass hierarchy) the relative 
size of the Earth matter
effects increases with energy and can be seen as oscillatory
modulation of the energy spectrum. For small energies, however, this
modulation oscillates too fast to be resolved, \ie, it averages out. For large
energies about $25 \, \mathrm{MeV}$, the frequency becomes smaller and this
modulation can be resolved. In \fig~\ref{Rates}, we used the scenario S1 from
\tab~\ref{snscenarios}, making it possible that the fluxes of $\nu_x$ dominate 
above the critical energy around $25 \, \mathrm{MeV}$. This
results in a negative modulation of the electron antineutrino spectrum, \ie,
regeneration effects of $\bar\nu_\mu$ and $\bar\nu_\tau$, as it is shown in
\Ref~\cite{Lunardini:2001pb}. However, a general suppression of the fluxes of
the $\nu_x$ is possible for a value $\xi<1$, which means that the
modulation can be positive. From \fig~\ref{Rates}, it is also interesting to
observe that solar neutrinos show much smaller Earth matter effects
(day-night), since the spectrum is cut off far below $25 \,
\mathrm{MeV}$. Thus, especially the high-energy
tail in the supernova neutrino spectrum can make this application
possible compared with solar neutrinos.

Let us now go back to \fig~\ref{fig:fr}, which shows the flux ratio $f_R$
of the $\nu_x$ and $\bar\nu_1$ fluxes at the surface of the Earth for
some of
the supernova parameter scenarios in \tab~\ref{snscenarios}. With
\eqs~(\ref{eq:FD}), (\ref{eq:deffr}), and (\ref{eq:fluxd2}) as well as
this figure, we can estimate the sensitivity for different sets of supernova
parameters. This equation describes the flux at the detector $D_2$ and depends
on three different quantities $\Phi_{\bar\nu_1}^0$, $f_R$, and
$P_{1e}^{D_2}$. First, the flux $\Phi_{\bar\nu_1}^0$ can be indirectly
determined by the detector $D_1$, since it also appears in
\eq~(\ref{eq:fluxd1}) as the enveloping function, \ie, both fluxes
$\Phi_{\bar \nu_e}^{D_1}$ and $\Phi_{\bar \nu_e}^{D_2}$ are directly
proportional to $\Phi_{\bar \nu_1}^0$. Second, in general, the 
flux ratio $f_R$ depends on the supernova parameters $\xi$ and
$\tau_E$ as well as the $\eta$'s, which can, up to a certain
precision, also 
be reconstructed from the spectrum at the detector $D_1$ (\cf, discussion in
\Sec~\ref{Sec:uncertainties} and \Refs~\cite{Barger:2001yx,Minakata:2001cd}). 
Note that it could be directly measured if one
were also able to detect flavors other than the electron antineutrino, and the
supernova parameters would completely drop out (\cf, \eq~(\ref{eq:fluxD1})). 
Third, the transition probability $P_{1e}^{D_2}$ contains the
information about the  Earth matter and is usually quite large. Thus,
the ratio $(1- P_{1e}^{D_2})/P_{1e}^{D_2}$ in \eq~(\ref{eq:fluxd2}) is
rather sensitive to changes in the Earth matter effects. Since this factor is
multiplied with $f_R$, the energy-dependent flux ratio can enhance or suppress
it. Finally, we have also noticed above that the (relative) Earth
matter effects are increasing with energy.
For the supernova parameters we can then distinguish four different 
cases, where some of those can also be found in \fig~\ref{fig:fr}:
\begin{enumerate}
\item $\xi=1$, $\tau_E=1$, $\eta_{\bar{\nu}_e}=\eta_{\bar{\nu}_\mu}=0$ 
(energy equipartition and equal temperatures for all flavors): The flux 
ratio $f_R$ is equal to unity (\cf, \eq~(\ref{eq:deffr})). Then, the 
neutrino transition probabilities in \eq~(\ref{eq:fluxd2}) drop out and 
we cannot use the supernova neutrinos for Earth matter effects.
\item
$\xi=1$, $\tau_E>1$, $\eta_{\bar{\nu}_e}=\eta_{\bar{\nu}_\mu}=0$ 
(energy equipartition and a lower temperature for
$\bar\nu_e$ than for $\nu_x$): The flux ratio $f_R$ is enhanced for large
energies, where Earth matter effects are large.
The larger $\tau_E$ is, the larger becomes this effect. Thus,
the scenario S2 in \tab~\ref{snscenarios} performs worse than the
scenario S1.
\item
$\xi \ll 1$ (more electron antineutrinos produced than the other two
flavors): The flux ratio $f_R\propto \xi$ is suppressed in
general. Therefore, the scenarios S3 and S4 are
not as good for our application as the scenario S1.
\item
$\tau_E>1$ and $\xi>1$ or $\eta_{\bar{\nu}_e} \neq
\eta_{\bar{\nu}_\mu}$ (more muon/tau antineutrinos produced than  
electron antineutrinos or different degeneracy parameters for the 
different flavors): The flux ratio $f_R$ becomes even steeper than the 
one for S1, reflecting the more different behavior of the different 
flavors. Thus, we expect that the scenarios S5 to S10 in 
\tab~\ref{snscenarios} perform much better than the
scenario S1, while the optimal scenario should be S9 having the largest 
$\xi$ and $\tau_E$.
\end{enumerate}
Thus, the larger $\xi$ and $\tau_E$ are and the more different 
$\eta_{\bar{\nu}_e}$ and $\eta_{\bar{\nu}_\mu}$ are, the better our 
proposed application should work. However, the actual values of these
parameters will not be known for sure before the next supernova explodes. 
Therefore, we choose further on the scenario S1 for reference, which 
has quite plausible parameter values and will turn out to be a
possible model that perfectly illustrates our procedure. 
In addition, we will summarize the performance 
of the other scenarios in short form.

For our application we assume that neutrino mass eigenstates arrive at the 
Earth and no
neutrino oscillations take place between the supernova envelope and the 
surface of the Earth. This can either be justified by the adiabacity condition 
for the
propagation within the supernova, making mass eigenstates emerge from it, or by
decoherence of neutrino oscillations between the surface of the supernova and
the Earth. In both cases, \eq~(\ref{eq:atotal}) can be split into two
independent factors without interference terms. The issue of wave packet
decoherence has, for example, been addressed in
\Refs~\cite{Giunti:1991ca,Giunti:1998wq,Grimus:1998uh,Cardall:1999ze,Lindner:2001th,Giunti:2002xg}. It has been found that neutrino
oscillations vanish for neutrino propagation over distances much
larger than the coherence length of the neutrino oscillations. This
means that for $L>L^{\mathrm{coh}}_{ab} \propto \sigma
E^2/\Delta m_{ab}^2$ the $L/E$-dependent interference terms produced
by the superposition of the mass eigenstates $m_a$ and $m_b$ in the
neutrino oscillation formulas are averaged out. The quantity $\sigma$
corresponds to a wave packet width determined by 
the production and detection processes~\cite{Giunti:1998wq,Cardall:1999ze}.
Since for supernova neutrinos the distance of the propagation is especially
large, it is plausible to assume that this averaging takes place and neutrino
oscillations vanish by natural decoherence. In other words, the different group
velocities of the wave packets of different mass eigenstates combine with
dispersive effects such that the overlap of the mass eigenstates is
gradually reduced to zero by a factor of $\exp \left( - 
[l/L^{\mathrm{coh}}_{ab}]^2 \right)$ in the neutrino oscillation formulas. 
Hence, for $L>L^{\mathrm{coh}}_{ab}$ the
mass eigenstates arrive separately and the coherent transition amplitude
in \eq~(\ref{eq:totalop}) is split up into two parts to be summed over 
incoherently (see, \eg, \Ref~\cite{Lindner:2001th}), having the same effect as
the adiabatic transition within the supernova separating the flavor states into
mass eigenstates. Thus, it is reasonable to assume that mass eigenstates arrive
at the surface of the Earth even for non-adiabatic transitions within the 
supernova.

Finally, in our numerical analysis, we assume Super-Kamiokande-like
water-Cherenkov 
detectors, \ie, a $32 \, \mathrm{kt}$ fiducial mass (for supernova neutrinos)
Super-Kamiokande detector and a  $1 \, \mathrm{Mt}$ fiducial mass (for
solar neutrinos) Hyper-Kamiokande detector. We choose 30 energy bins
between the threshold energy $5 \, \mathrm{MeV}$ and $70 \,
\mathrm{MeV}$, since above $70 \, \mathrm{MeV}$ the number of events
is rather low.
Furthermore, we assume a realistic energy resolution implemented with
Gaussian energy smearing with a smearing width proportional to
$\sqrt{E}$ such that we have an energy resolution of about 15~\% at $10
\, \mathrm{MeV}$ \cite{Bahcall:1997ha}.

\section{A neutrino oscillation tomography model}

\begin{figure}[ht!]
\begin{center}
\includegraphics*[width=7cm]{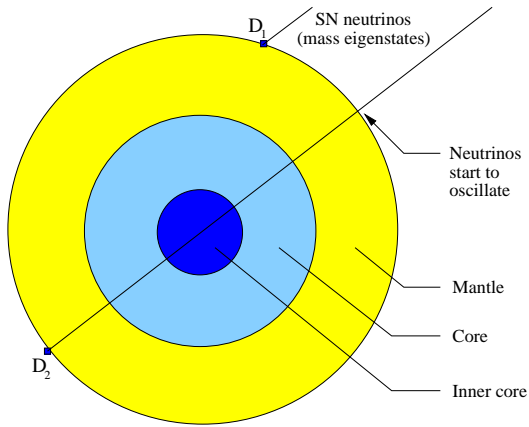}
\end{center}
\vspace{-1cm}
\caption{The minimal required setup for supernova
neutrino tomography with at least two baselines, one of which is ending
at the surface of the Earth at the detector $D_1$, the other one is going
through the Earth's core (or even the Earth's inner core) to the detector
$D_2$. In general, we assume that the neutrinos arrive as mass
eigenstates at the detector $D_1$ and start to oscillate when they
enter the Earth's interior.}
\label{SNbaseline}
\end{figure}

We now introduce and discuss a simple model used for supernova
neutrino tomography. As shown in \fig~\ref{SNbaseline},
we assume at least two baselines ending at detectors with similar
statistics and systematics, such as Super-Kamiokande-like
water-Cherenkov detectors.
In order to
measure the reference spectrum of the supernova neutrinos, the neutrinos
detected at the detector $D_1$ must not cross the Earth. 
If we want to obtain information on the Earth's
core, then the second detector $D_2$ needs to observe the supernova
neutrinos with a baseline crossing the Earth's core with a sufficient
length.
The probability to have such a
  configuration depends upon the location of the supernova in our Galaxy,
  the time of the day at which the burst arrives at the Earth, and the
  position of the detectors $D_1$ and $D_2$ themselves. For example,
  if we consider 
  a supernova located in the galactic center and the detectors placed in
  Japan and the USA\footnote[7]{For instance, one possible site for the
  UNO proposal would be the WIPP in Carlsbad, New
  Mexico, USA~\cite{UNO}.}, then, following
  \Ref~\cite{Lunardini:2001pb}, one obtains that the probability to find the
  required setup is around 25~\% and 10~\% along a day for
  neutrinos crossing the core and the inner core, respectively.
  Thus, although this configuration would not work out for many
  locations of a supernova in the galactic plane, it would be quite
  likely for a supernova being in the region of the galactic center.

Further on, we assume that the detector $D_1$ is at least as good as
the detector $D_2$, which means that the statistics is limited by
$D_2$, and $D_1$ measures the reference flux $\Phi_{\bar\nu_e}^{D_1}$ with
sufficient precision. For the neutrino oscillation parameters we choose
$\theta_{12}=32.9^\circ$, $\theta_{13}=5^\circ$,
$\theta_{23}=45^\circ$, $\Delta m_{21}^2 = 5.0 \cdot 10^{-5} \,
\mathrm{eV}^2$, $\Delta m_{32}^2 = 2.5 \cdot 10^{-3} \,
\mathrm{eV}^2$, and $\delta_{\mathrm{CP}}=0$
\cite{Bahcall:2002hv,Apollonio:1998xe,Apollonio:1999ae,Bemporad:1999de,Toshito:2001dk},
\ie, a 
normal mass hierarchy. Matter effects on supernova neutrinos in the Earth are
discussed in detail in \Ref~\cite{Lunardini:2001pb}, where it is also
demonstrated that Earth matter effects would be suppressed for solutions other
than the MSW LMA solution.
In addition, for detecting antineutrinos
the Earth matter effects would only be large enough for our
application with an inverted mass hierarchy if $|U_{e3}|^2$ is
larger than about $10^{-5}$ \cite{Lunardini:2001pb}. Thus, we specialize on the
(not unlikely) case of the MSW LMA solution with a normal mass
hierarchy in order to be able to observe Earth matter effects with a
sufficient precision.
For our application the dominant neutrino
oscillation parameters are the solar neutrino oscillation parameters
$\Delta m_{21}^2$ and $\theta_{12}$, as
well as the matter effects depend on $\sin^2 2 \theta_{13}$. Later, we will
estimate the precision with which we need to know these parameters and
we will test the influence of the size of $\sin^2 2
\theta_{13}$. Note that we assume mass eigenstates arriving at the
surface of the Earth. Therefore, if $D_1$ and $D_2$ were
identical detectors, a direct comparison of their energy spectra would verify
the existence of Earth matter effects immediately. For detectors of
different types fits of the energy spectra would supply similar information in
an indirect way.

\begin{figure}[ht!]
\begin{center}
\includegraphics*[width=7cm]{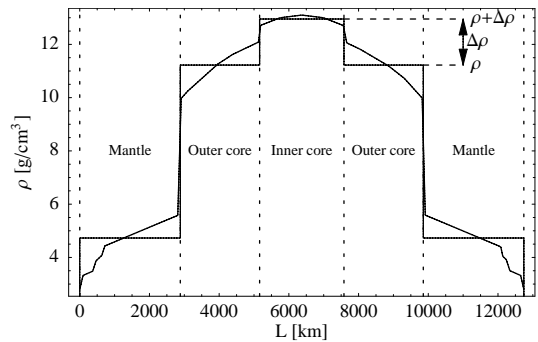}
\end{center}
\vspace{-1cm}
\caption{The model for the Earth's matter density
profile used in the calculations (step function) and the PREM
matter density profile as function of the path length along the baseline
shown in \fig~\ref{SNbaseline}. The quantities which we are interested
in are the average outer core matter density $\rho$ and the matter density jump
$\Delta \rho$ between average inner and outer core densities, as it is
shown in this figure.}
\label{SNProfile}
\end{figure}

For the modeling of the matter density profile we choose the
Preliminary Reference Earth Model (PREM) profile \cite{stac77,dzie81}, as
it is shown in \fig~\ref{SNProfile},
and approximate it by layers of constant average matter
densities. A baseline with a maximum length of twice the Earth radius then
crosses the following layers in this order: mantle, outer core, inner
core, outer core, mantle. Since substantial knowledge is provided by geophysics
about the Earth's mantle, we assume its matter density to be known with a
sufficient precision. For such a baseline the interesting quantities
to measure are the average outer core matter density $\rho$ and the
matter density jump $\Delta \rho$ to the average inner core matter
density, as illustrated in \fig~\ref{SNProfile}. This is slightly 
different to what is known from seismic wave geophysics, since there
the density jumps of the actual matter densities at the mantle-core and
outer-inner core boundaries are better known. However, since neutrino
oscillations are not sensitive to matter densities at individual points,
but essentially to the integral of the matter density and the length
scale the neutrinos are traveling through~\cite{Jacobsson:2001zk}, they are 
more appropriate to measure average matter densities.

The introduced model allows to estimate
what could be learned from the neutrinos of a supernova explosion 
about the Earth's interior. The actual situation, however, such as
the number of detectors with baselines through the Earth, their size, the
knowledge on the neutrino oscillation parameters, systematics, \etc, can only 
be implemented after the next observed supernova explosion. Our
discussion here serves only the purpose of demonstrating that such
studies are feasible.

\section{Results}

Based on the modeling in the last section, we present results, which
could be provided by a single supernova. Our analysis is
performed with a standard $\chi^2$ technique using~\cite{Groom:2000in}
\begin{equation}
\chi^2 \equiv 2 \, \sum_{i=1}^n \left( x^{\mathrm{ref}}_i - x_i + x_i \,
\mathrm{log} \frac{x_i}{x_i^{\mathrm{ref}}} \right),
\end{equation}
where $n$ is the number of energy bins, $x_i^{\mathrm{ref}}$ is the
reference event rate in the $i$th bin for the true parameters, and
$x_i$ is the measured/varied event rate in the $i$th bin. The errors quoted 
are read off at the $2 \sigma$ confidence level -- depending on the problem 
for one or two degrees of freedom. For two degrees of freedom we also take into
account the two-parameter correlations. However, we assume that the
effects of the systematics are negligible, \ie, the systematical errors are not
larger than the statistical errors and the systematics is well understood. This
assumption should be reasonable at the time when this application
could become relevant.

The most likely case to observe the Earth's core with supernova neutrinos
are baselines crossing the Earth's outer core, but not the Earth's inner core, 
\ie, baselines between about $10{,}670 \, \mathrm{km}$ and $12{,}510 \, 
\mathrm{km}$.
Assuming the mantle properties
to be known quite well from geophysics, one may then measure the
average (outer) core matter density (about $11.4 \,
\mathrm{g/cm^3}$). 
As a result of the analysis for the scenario S1 in
\tab~\ref{snscenarios}, it turns out that one could measure this core
matter density with a baseline just touching the inner core ($L \simeq
12{,}510 \, \mathrm{km}$) with about 9~\% precision with a
Super-Kamiokande-like detector and 1.3~\% precision with a
Hyper-Kamiokande-like detector. We find a rather small dependence on
the baseline as long as $L \gtrsim 11{,}250 \, \mathrm{km}$. For $L
\simeq 11{,}250 \, \mathrm{km}$ we still find precisions of 16.5~\%
(Super-Kamiokande) and 2.5~\% (Hyper-Kamiokande), which are growing
with an increasing baseline length. The reason for the weak baseline
dependence comes from the geometry of the problem, which demonstrates
that above a certain threshold the traversed core fraction is rather
large, but hardly changes if the total baseline length is further
increased. Thus, we will further on not discuss the baseline
dependence, since it turns out to be negligible close to the maximal
traversed core/inner core distance.

\begin{figure*}[ht!]
\begin{center}
\includegraphics*[width=15cm]{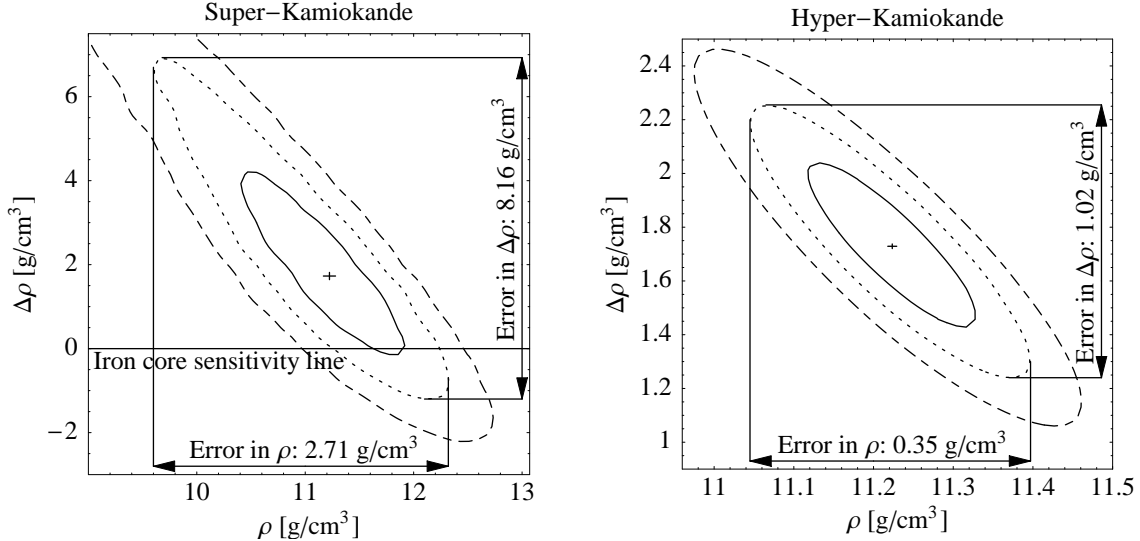}
\end{center}
\vspace{-1cm}
\caption{The $1 \sigma$, $2 \sigma$, and $3 \sigma$
contours of the $\chi^2$-function for a measurement of $\rho$ and $\Delta \rho$
for a Super-Kamiokande-like (left-hand plot) and Hyper-Kamiokande-like
(right-hand plot) detector and the scenario S1 in
\tab~\ref{snscenarios}
without uncertainties and systematical errors. For the detectors, we
use the ones described at the end of \Sec~\ref{Sec:earth}.
The
errors from statistics and correlations are read off at the arrows in the
figures. In order to find evidence for the existence of the inner core (the
iron core),
the contour of the respective confidence level must not cut the
inner core sensitivity line corresponding to $\Delta \rho \equiv 0$.
This line can only be seen in the left-hand plot. In addition, in the
left-hand plot, degenerate solutions are, in principle, present at the
$2\sigma$ confidence level, which are, however, out of the plot range.}
\label{SNRes}
\end{figure*}

A somewhat more sophisticated application is the combined measurement of
the outer and inner core matter densities in the two-parameter model 
introduced in the
last section, \ie, \figs~\ref{SNbaseline} and~\ref{SNProfile}. In
\fig~\ref{SNRes},
the results of this analysis without systematical
errors and uncertainties are shown
in the $\rho$-$\Delta
\rho$-plane for Super-Kamiokande- and Hyper-Kamiokande-like detectors for the
scenario S1 in \tab~\ref{snscenarios}.
One can read off the (absolute) errors at the $2 \sigma$ confidence level such
as shown in this figure. A Super-Kamiokande-like
detector cannot verify the inner core, which can be seen in the
contours crossing the inner core sensitivity line, \ie, the line $\Delta \rho
\equiv 0$. Furthermore, the $3 \sigma$ contour is rather extensive and
cut-off in the upper left corner.
Therefore, it is not well-suited for density measurements
of the inner core. However, a Hyper-Kamiokande-like detector can
clearly observe and verify the inner core even at the 99~\% confidence
level. Furthermore, quite precise measurements of $\rho$ and $\Delta
\rho$ are possible. The relative error for $\rho$ is about 3.1~\% and
for $\Delta \rho$ about 59~\% ($2 \sigma$ confidence level).

\begin{table*}
\begin{center}
\begin{tabular}{lccccrrc}
\hline
Scenario & $\xi$ & $\tau_E$ & $\eta_{\bar{\nu}_e}$ 
&$\eta_{\bar{\nu}_\mu}$ & $\delta \rho$ & $\delta (\Delta \rho)$ & Degs \\
\hline
S1 & 1.0 & 1.4 & 0 & 0 & 2.7 & 8.2 & Yes \\
S2 & 1.0 & 1.2 & 0 & 0 & 6.7 & 17.5 & Yes \\
S3 & 0.5 & 1.4 & 0 & 0 & $\gtrsim 8$ & $\gtrsim 21$ & Yes \\
S4 & 0.5 & 1.2 & 0 & 0 & $\gtrsim 12$ & $\gtrsim 30$ & Yes \\
S5 & 1.2 & 1.4 & 0 & 0 & 2.3 & 7.0 & No \\
S6 & 1.2 & 1.2 & 0 & 0 & 5.0 & 12.5 & Yes \\
S7 & 1.0 & 1.4 & 3 & 1 & 2.0 & 5.9 & No \\
S8 & 1.0 & 1.2 & 3 & 1 & 3.7 & 10.3 & Yes \\
S9 & 1.2 & 1.4 & 3 & 1 & 1.7 & 5.3 & No \\
S10 & 1.2 & 1.2 & 3 & 1 & 3.1 & 9.1 & Yes \\
\hline
\end{tabular}
\end{center}
\caption{The different supernova parameter scenarios from
\tab~\ref{snscenarios} and the absolute errors $\delta \rho$  and $\delta
(\Delta \rho)$ (in $\mathrm{g/cm^3}$) for the measurement of $\rho$ and 
$\Delta
\rho$, respectively, at the $2 \sigma$ confidence level with a
Super-Kamiokande-like detector. In addition, the appearance of degenerate
solutions (Degs) at the $2 \sigma$ confidence level is indicated (for 
Super-Kamiokande).}
\label{scresults}
\end{table*}
As we have indicated in \Sec~\ref{Sec:earth}, the supernova parameter scenarios
in \tab~\ref{snscenarios} other than the scenario S1 perform somewhat worse in 
the measurement of the parameters $\rho$ and $\Delta \rho$. In addition,
degenerate solutions appear at the $2 \sigma$ confidence level, \ie, different 
solutions in the parameter space can be fit to the results of the measurement 
at the considered confidence level. In
\tab~\ref{scresults}, we show the errors for the measurement of $\rho$
and $\Delta \rho$ for the Super-Kamiokande detector.
It clearly reflects the behavior expected in \Sec~\ref{Sec:earth} 
and it demonstrates that large values of $\xi$ and $\tau_E$ as well as 
different values for the $\eta$'s of the different flavors give the best 
results. It also shows that the scenario S1 is one of the the most
conservative choices of the scenarios for which the tomography
application would work.
In addition, note that degenerate solutions are in many solutions 
present due to the energy resolution function of the detector.
It turns out that a somewhat higher $\langle E_{\bar\nu_e} \rangle$
can improve the results, because it supports the high-energy tail in
the spectrum where matter effects are largest.

The results from this measurement cannot be directly compared with the 
geophysical results, because
in seismic wave geophysics the matter density jumps are much easier
accessible than the average matter densities. For example, the matter
density jump between the outer and inner cores is believed to be about
$(0.55 \pm 0.05) \, \mathrm{g/cm^3}$ (see, \eg,
\Refs~\cite{MastersShearer:90,ShearerMasters:90,IshiiTromp:99,MastersLaskeGilbert:00}). Translated to the $2 \sigma$ confidence level, this
corresponds to the same order of magnitude as the Hyper-Kamiokande measurement
of about 59~\% precision. However, the average matter density is much
harder to access in geophysics and can only be estimated by the long
periodic seismic eigenmodes with uncertainties increasing with 
depth~\cite{Igel}. The precision on the average matter
density of about 3~\% from neutrino physics as well as the measurement of the
average matter density jump $\Delta \rho$ could help to understand and
complement the geophysical information.

Finally, one could consider more than one
baseline. If more than one detector observes a supernova
through the Earth's core or the inner core, then the potential of this
technique would be improved. However,
the probability for a single detector to
have a baseline through the core is already quite low, which means
that more detectors would mainly increase the probability
that one has an appropriate baseline. Thus, we have in this paper
focused on the case of one baseline through the Earth. If really more
than one baseline went through the core, then the statistics of the overall
measurement would be improved and the result could be estimated by a
scaling of the detector. Having one large
detector and one baseline corresponds in this application to different
detectors at similar positions with their fiducial masses adding up to the one
of the single detector.

\section{Uncertainties}
\label{Sec:uncertainties}

So far, we have taken into account for our analysis only the
statistical errors and the larger two-parameter-correlation.
Once illustrated the advantages of our procedure,
we now discuss the influence of uncertainties on the measurements of
the parameters. We do this separately, because the sensitivity to the parameter
extraction, in particular, to the astrophysical parameters, depends
significantly upon the model considered.

In order to discuss the influence of uncertainties on the
measurements, we estimate the precision to which the leading neutrino
oscillation parameters  $\Delta m_{21}^2$ and $\theta_{12}$ have to be
known for the measurement and we vary them
until we observe an effect which is as large as the error of the measurement of
$\rho$ or $\Delta \rho$. It turns out that these leading parameters have to be
known with about 1~\% precision for the Super-Kamiokande-like measurements and
with about 0.2~\% precision for the Hyper-Kamiokande-like measurements. These
precisions should be obtainable on the typical timescales of galactic 
supernova explosions.

The parameter $\sin^2 2 \theta_{13}$ also has some influence on the matter
effects. For small values, however, the (three-flavor) neutrino
oscillations reduce to two two-flavor neutrino oscillation schemes one
describing the solar neutrino oscillations and the other one the
atmospheric neutrino oscillations. We tested the influence of this parameter on
our applications and we found that it can be safely neglected as long as 
$\sin^2 2 \theta_{13}$ is not too large. Only at the CHOOZ bound minor 
corrections much smaller than the error of our measurements have to be
performed. However, this bound will be reduced in the short term
future by planned superbeam and neutrino factory experiments (for a
summary of expected boundaries see, for example, \Ref~\cite{Huber:2002mx}).

One of the most important uncertainties in this measurement comes from
measuring the electron antineutrinos only. It can be easily seen from
\eq~(\ref{eq:fluxd1}) that the extraction of the fluxes
of the mass eigenstates from the flux of the electron antineutrino flavor 
involves
assumptions about the supernova parameters, which are entering by the flux
ratio $f_R$. Thus, if only the electron antineutrinos from the supernova can be
measured, then the tomography problem will be closely connected to the
determination of the supernova parameters at a detector on the surface
of the Earth.
\begin{table*} \begin{center}
\begin{tabular}{lccc}
 \hline
 Param. & SuperK & HyperK & Effects  \\
 \hline
$E_{\bar{\nu}_e}^{\mathrm{tot}}$ & $\sim 5~\%$ & $\sim 1~\% - 2~\%$ &
 $\lesssim 1~\%$ \\
$E_{\bar{\nu}_\mu}^{\mathrm{tot}}$ & $ \sim 100~\%$ & $\sim 10~\%$ & large \\
$\langle E_{\bar{\nu}_e} \rangle$ & $ \sim 5~\%$ & $\sim 1~\% - 2~\%$ &
$\lesssim 5~\% - 10~\%$ \\
$\langle E_{\bar{\nu}_\mu} \rangle$ & $ \sim 10~\%$ & $\sim 1~\% -
 2~\%$ & $\lesssim 5~\%$ \\
 \hline
\end{tabular}
\end{center}
\caption{The uncertainties on the supernova parameters
extracted from the Super-Kamiokande (SuperK) or Hyper-Kamiokande
(HyperK) measurements from \Refs~\cite{Barger:2001yx,Minakata:2001cd} and their
estimated effects on our parameter measurements as percentage corrections.}
\label{snparams}
\end{table*}
Estimates for the precision of the supernova
parameter determinations depend on the initial values
considered. Thus, the more different the $\bar\nu_e$ and $\bar\nu_\mu$
spectra are formed in a supernova explosion, the stronger the
oscillation effects are, and therefore, the better one can measure the
neutrino temperatures and luminosities. As an example, we show in
\tab~\ref{snparams} the sensitivities obtained in
\Refs~\cite{Barger:2001yx,Minakata:2001cd}.\footnote[8]{The results in 
this reference were obtained for $\eta_{\bar\nu_e} = 2.6$,
$\eta_{\bar\nu_\mu} = 0$, and $\xi=0.5$. In this paper, however, we
have used slightly different parameter values for which the precisions
on the supernova parameters in \tab~\ref{snparams} would become
somewhat worse by about a factor of two.} In addition, this table
shows the estimated influences on the determination of our parameters
from a variation of the supernova parameters in the numerical
evaluation. The reason for the similar percentage effects for
Super-Kamiokande and Hyper-Kamiokande is the parallel 
scaling of both problems. It is interesting to observe that none of the
supernova parameters has a strong influence on the tomography problem except 
from the overall energy of the muon antineutrinos
$E_{\bar{\nu}_\mu}^{\mathrm{tot}}$. One can show that this
parameter has to be known up to about 20~\% for the Super-Kamiokande
measurement and 3~\% - 4~\% for 
the Hyper-Kamiokande measurement in order not to have strong effects on the
tomography problem. However, this precision cannot be achieved by measuring the
electron flavor only.
Altogether, either $E_{\bar{\nu}_\mu}^{\mathrm{tot}}$
needs to be measured by different experiments or the fluxes of the muon
and tau antineutrinos need to be determined simultaneously with the electron
antineutrino flux, making the supernova parameters entirely drop out. This can
be seen in \eq~(\ref{eq:fluxD1}), which allows the reconstruction of all mass
fluxes at the detector $D_1$ if all flavor fluxes are measured.

Another issue in the discussion of uncertainties is the parameter $Y_e$ in
\eq~(\ref{eq:A}) relating the number of electrons to
the number of nucleons.
Since the Earth matter effects in neutrino oscillations actually 
depend on the electron density and not on the matter density directly,
additional uncertainties enter in the conversion of these two quantities by the
parameter $Y_e$. We assumed $Y_e=0.5$ in our calculations, but for different
materials this parameter can differ somewhat from this value -- especially
in the inner core. In order to find out the material in each matter density 
layer, one
may prefer to measure the electron density instead of the matter density.
However, since in each layer these quantities are proportional to each other,
the problem does not change by using the matter density and the
electron density can be easily calculated.

So far, we have assumed the average mantle matter density to be
known exactly. In fact, matter density uncertainties up to
5~\% have been documented (for a summary, see, \eg,
\Ref~\cite{Panasyuk}). Testing these uncertainties, even the
conservative choice 
of a 5~\% uncertainty on the average mantle density is no
problem for the Super-Kamiokande detector. However, for the
Hyper-Kamiokande detector, it should not be larger than about
2~\%. Because of a partial averaging out of the uncertainties for the
very long baselines considered in this paper, this uncertainty should
be quite realistic~\cite{Jacobsson:2001zk}.

Another source of uncertainty
has to do with the knowledge of the location of a supernova, since it
may affect the baseline length. In a future core-collapse supernova in
our Galaxy, the electromagnetic radiation can be obscured by dust in
the interstellar space. In this case, the localization of the supernova can
be done by studying neutrino-electron forward scattering. A
conservative estimate leads to restrict the supernova direction
with an error of about $5^\circ$ in the Super-Kamiokande
detector~\cite{Beacom:1998fj}.
However, with the increase of statistics expected in a megaton detector,
this uncertainty would be reduced. On the other hand, complementary information
could be inferred by the triangulation method \cite{Beacom:1998fj}, namely
the direct measurement of the delay of the signal to the two detectors.
Therefore, we consider that this small error would not imply a significant
change in the long baselines that we are dealing with, since even for
a $5^\circ$ directional uncertainty the baseline length of about
$12{,}700 \, \mathrm{km}$ does not change very much.

\section{Summary and conclusions}
\label{sec:S&C}

We have discussed the possibility to use the neutrinos from a future galactic
supernova explosion to obtain additional information on the Earth's
core. First, we have summarized geophysical aspects and unknowns of the
Earth's core. Then, we have investigated core-collapse supernovae as
potential neutrino sources for a snapshot of the Earth's
interior. Next, we have discussed the neutrino propagation from the
production to the detection in detail, where we have especially focused
on Earth matter effects on the neutrino oscillations of the supernova
neutrinos. We have also put these effects into the context of the
supernova parameters, \ie, temperatures and deviation from energy 
equipartition. Furthermore, we have stressed the importance of supernova
neutrinos arriving at the surface of the Earth as mass eigenstates for
this technique, which we have also supported by a discussion of decoherence of
neutrino oscillations. We have shown that we need one detector on the surface 
of the Earth on the side towards the supernova, and another one in the shadow 
of the Earth's core. For the most likely scenario of not crossing the Earth's 
inner core, we have shown that the
Earth's average core matter density could be determined up to 9~\% with a 
Super-Kamiokande-like and 1.3~\% with a Hyper-Kamiokande-like
detector (all errors at the $2 \sigma$ confidence level). 
In addition, for a less likely two-parameter measurement of the outer
and inner core matter densities, Hyper-Kamiokande could verify the existence
of the inner core at the $3 \sigma$ confidence level and measure the outer
core matter density with a precision of about 3.1~\%, as well as the
density jump
between outer and inner core matter densities with a precision of
about 59~\%. The latter error is comparable to seismic wave geophysics, where, 
however, not
the difference between the average matter densities, but the matter density 
jump at
the outer-inner core boundary is measured. Thus, neutrino physics
could provide complementary information to geophysics. 
However, the quoted numbers for the precisions depend on the
supernova parameter values, indirectly determined by the on-the-surface
measurement, and are in some cases better, in others worse. Thus, the
actual precisions will not be known before the supernova goes off.
In general, we find that the more muon and tau neutrinos are produced, 
the larger the temperature difference between the different flavors is, 
and the more different the degeneracy parameters for the different 
flavors are, the better the application works.
Finally, we
have discussed several uncertainties to these measurements and we have found 
that especially the determination of the total muon antineutrino energy of
the supernova causes
problems to our method. However, measuring not only electron antineutrinos, but
also the other two flavors could completely eliminate the dependence
on the supernova parameters. Furthermore, the leading solar neutrino parameters
have to be known with sufficient precision, which is about 0.2~\%
for Hyper-Kamiokande-like measurements. In summary, supernova neutrino
tomography could be a nice additional payoff of existing or planned detectors 
if all of the prerequisites can be met at the time when the next
supernova explodes.

\section*{Acknowledgments}

We would like to thank Heiner Igel for useful discussions and
Evgeny~Kh.~Akhmedov and John~F.~Beacom for valuable comments.

This work was supported by the Swedish Foundation for International
Cooperation in Research and Higher Education (STINT) [T.O.], the
Wenner-Gren Foundations [T.O.], the Swedish Research Council
(Vetenskapsr{\aa}det), Contract No.~621-2001-1611 [T.O.], the Magnus
Bergvall Foundation (Magn.~Bergvalls Stiftelse) [T.O.], the ``Deutsche
Forschungsgemeinschaft'' (DFG) [R.T.], the
``Studienstiftung des deutschen Volkes'' (German National Merit
Foundation) [W.W.], and the ``Sonderforschungsbereich 375 f{\"u}r
Astro-Teilchenphysik der Deutschen Forschungsgemeinschaft'' [M.L.,
  T.O., and W.W.].

\end{document}